# Thermodynamics Of Finite Quantum Systems: Application To Spin Magnetism II

S. Sitotaw and R.A. Serota

Department of Physics

University of Cincinnati

Cincinnati, OH 45221-0011

Abstract

We extend our study of thermodynamics of a Kubo particle to temperatures smaller than the mean interlevel spacing. We obtain the distribution functions of spin susceptibility and heat capacity for Poisson and Wigner-Dyson level statistics. We evaluate the line shape of the Knight shift due to spin effects both in a single particle and for the ensemble average and compare it with orbital and spin-orbit contributions.

1.      The spin susceptibility of an ensemble of small particles with Poisson level statistics was first evaluated by Kubo [1]. Building on the absence of level correlations, he found the exact result both for the odd and even electron particles. Gor'kov and Eliashberg argued [2] that disordered particles and particles with diffusive boundaries should have Wigner-Dyson (WD) level statistics [3]. Denton, Mühschlegel, and Scalapino [4] have noticed that Kubo's results follow from a two- and three-level model in the even and odd cases respectively and evaluated the electron spin susceptibility and heat capacity for the WD circumstance. These developments have been extensively reviewed [5].

2.      Despite the realization of the statistical nature of spin response, the authors of these works have been concerned only with the average susceptibility. On the other hand, the test of their predictions relied mainly on the measurements of the Knight shift [6]. Obviously, the latter should be broadened due to the distribution of spin susceptibilities. Although the inhomogeneous broadening has been of concern [5], it was attributed largely to the particle size distribution. Below we obtain the distribution of the spin susceptibility using the model of Ref. [4] and discuss its effect on the inhomogeneous broadening of the Knight shift vis-a-vis the fluctuations of the wave function in the particle [7]. We compare this effect to the inhomogeneous broadening due to orbital and spin-orbit effects [8]. We also evaluate the distribution function of the heat capacity.

3.      In what follows, we set the Curie susceptibility, $\chi_C = \mu_B^2 \beta$, where $\beta = \left( k_B T \right)^{-1}$ and $\mu_B$ is the Bohr magneton, to unity. In these units, the Pauli susceptibility is $\chi_P = \varepsilon$, where $\varepsilon^{-1} \equiv \beta \delta \gg 1$ and $\delta$ is the mean interlevel spacing. For even electron particles, the partition function of the two-level system - the Fermi level, taken at zero, and the nearest unoccupied level at a distance $\Delta$ - in the magnetic field is given by [4], [5],

$$Z = 1 + 2\left( 1 + \cosh\left( 2\beta\mu_B H \right) \right)\exp\left( -\beta\Delta \right) + \exp\left( -2\beta\Delta \right),  \tag{1}$$

The distribution function for the susceptibility is obtained via integration of the Dirac delta-function whose argument is obtained from Eq. (1),

$$p\left(\chi\right) = \int_0^\infty d\Delta\, \delta\left(\chi - \frac{8t}{1+4t+t^2}\right) P\left(\Delta\right), \qquad t = \exp\left(-\beta\Delta\right), \tag{2}$$

where $P\left(\Delta\right)$ is the nearest level distribution function given by [3], [5],

$$P\left(\Delta\right) = \frac{1}{\delta}\exp\left(-\frac{\Delta}{\delta}\right), \qquad \text{Poisson}, \tag{3a}$$

$$P\left(\Delta\right) = \frac{\pi}{2}\frac{\Delta}{\delta^2}\exp\left(-\frac{\pi}{4}\left(\frac{\Delta}{\delta}\right)^2\right), \quad \text{orthogonal}, \tag{3b}$$

for the two cases considered here.

4.    The integral (2) can be trivially evaluated with the help of the formula,

$$\delta\left(f\left(x\right)\right) = \sum_i \left| f'\left(X_i\right)\right|^{-1} \delta\left(x - X_i\right), \tag{4}$$

where $X_i$ are the roots of the equation $f\left(X_i\right) = 0$. The result of exact integration, however, is quite messy as well as partially superfluous. A simplified expression can be obtained on the basis of the following argument. The mean susceptibilities are $\left\langle \chi \right\rangle \sim \varepsilon$ and $\left\langle \chi \right\rangle \sim \varepsilon^2$ in the Poisson and orthogonal cases respectively. Using the result of exact integration, it can be also shown that $\int_{\langle\chi\rangle}^{x_{\max}} d\chi\, p\left(\chi\right)$ is $\sim \varepsilon$ and $\sim \varepsilon^2$ respectively, while $\int_0^{\langle\chi\rangle} d\chi\, p\left(\chi\right) \sim 1$ in either case, where $\chi_{\max} \sim 1$ is the maximum value of the susceptibility found from Eqs. (2) and (4). We further notice that the accuracy of the distributions obtained from the two-level model is reduced near $\chi_{\max}$. Indeed, the $n$'th moments scale as $\varepsilon$ and $\varepsilon^2$ regardless of $n$ (see Appendix I), meaning that $\left\langle \left(\chi - \left\langle\chi\right\rangle\right)^n \right\rangle \sim \chi_C^{n-1}\chi_P$ in the Poisson and $\sim \chi_C^{n-2}\chi_P^2$ in the orthogonal cases respectively which, functionally, can be understood as $\chi_C^n$ multiplied by the probability of finding the next level within $k_B T$ of the Fermi level. The first of these relationships is consistent (see Appendix II) with the formal extension of our derivation [9] for $\beta\delta \ll 1$ to $\beta\delta \gg 1$. It indicates that the tails of the distributions are defined by the higher moments. However, as shown in Appendix I, such moments are not accurately defined by the two-level model, even for $\beta\delta \gg 1$.

5.      We conclude that whereas the distributions have long tails upward $\chi \sim 1$, the two-level model is insufficient for their accurate description. Since also the susceptibilities larger than the mean have a small probability, a good approximation is achieved by taking $t \ll 1$ in Eq. (2). Subsequent verification confirms that the expressions thus obtained,

$$p\left(\chi\right) = \frac{1}{8}\varepsilon\left(\frac{1}{8}\chi\right)^{\varepsilon - 1}, \qquad\qquad\qquad \text{Poisson,} \qquad\qquad (5a)$$

$$p\left(\chi\right) = -\frac{\pi\varepsilon^2}{2\chi}\exp\left(-\frac{1}{4}\pi\left(\varepsilon\ln\left(\frac{1}{8}\chi\right)\right)^2\right)\ln\left(\frac{1}{8}\chi\right), \qquad \text{orthogonal.} \qquad (5b)$$

coincide, within the accuracy of the model, with the result of exact integration (see Fig. 1). Both expressions have integrable singularities at $\chi \rightarrow 0$. Clearly, the small susceptibilities reflect the energy gap at the Fermi level.

6.      For odd number of electrons, one must consider the three-level model and $\langle\chi\rangle \cong 1$, irrespective of the level statistics [1], [4], [5]. From the latter fact alone one concludes that the distribution function is narrower than for the even electron circumstance. Formally, the susceptibility distribution function can be derived by averaging the delta function with the distribution function $P\left(\Delta, \Delta'\right)$ for the two levels [4]. We notice, however, that in the same approximation as above one finds the functions that diverge at $\sim 1$ similarly to Eqs. (5) at zero, with the extra factors of $\varepsilon$ and $\varepsilon^2$ respectively (see Appendix III) that can be traced to the probability of finding a level within $k_B T$ *below*, as well as above, the Fermi level. The distribution tails extend to the values of the susceptibility of order of the mean.

7.      The distribution function of the heat capacity has the same functional form for even and odd number of electrons and differs only in numerical coefficients. Consequently, we limit our discussion to the even case here. From the partition function in Eq. (2) for $H = 0$, we find for the distribution function,

$$p\left(C\right) = \int_0^\infty d\Delta \, \delta\left(C - \left(\beta\Delta\right)^2 \frac{4t}{1 + 4t + t^2}\right) P\left(\Delta\right), \qquad t = \exp\left(-\beta\Delta\right),\qquad(6)$$

where $C$ is measured in units of $k_B$. Making the same approximation as before, we have,

$$p\left(y\right) \approx \varepsilon^2 \int_0^\infty d\Delta \, \delta\left(y - 4\exp\left(-\beta\Delta\right)\right) P\left(\Delta\right)\qquad(6a)$$

where $y = C\varepsilon^2$. As a result, replacing $8 \rightarrow 4$, one finds the same expressions for $p\left(y\right)\varepsilon^{-2}$ as for $p\left(\chi\right)$ in Eqs. (5). Therefore, the heat capacity distributions have tails toward $C \sim 1$, while the mean values are $\sim \varepsilon$ and $\sim \varepsilon^2$ for the Poisson and orthogonal cases respectively.

8.     To test the predictions for the distributions of spin susceptibility and heat capacity experimentally, one must be able to measure the respective quantities in a single system. The rings and grains studied in recent experiments on orbital magnetic response [10] are a candidate but they are perhaps too large to achieve the condition $\left(\beta\delta\right)^{-1} \ll 1$ and perform a reliable measurement. The ensembles of particles studied earlier [5], as well as ensembles of rings [10], would yield only average quantities. The Knight shift broadening, however, would reflect the susceptibility distribution. For this reason, we now turn to its discussion in the light of the above results. As is well known [11], an electron creates a magnetic field at the nucleus site $\vec{r}$,

$$\vec{\mathcal{H}}\left(\vec{r}\right) = \frac{8\pi\mu_B}{3} \vec{\bar{\sigma}} \left|\psi\left(\vec{r}\right)\right|^2,\qquad(7)$$

where the over-bar denotes the quantum-mechanical average of the spin state. Performing now the thermodynamic averaging, we notice that in the limit $\left(\beta\delta\right)^{-1} \ll 1$ the electrons that occupy the levels below the Fermi level do not contribute (for the even electron case) since we neglect their excitations and their spins are pair-wise opposite. The two electrons at the Fermi level are allowed to jump to the next level in the two-level model. Of the six possible states, only two are magnetic yielding,

$$\mathcal{H}\left(\vec{r}\right) = \frac{8\pi\mu_B}{3} \frac{\left(\exp\left(-\beta\left(\Delta - 2\mu_B H\right)\right) - \exp\left(-\beta\left(\Delta + 2\mu_B H\right)\right)\right)}{1 + 4\exp\left(-\beta\Delta\right) + \exp\left(-2\beta\Delta\right)} \left(\left|\psi_0\left(\vec{r}\right)\right|^2 + \left|\psi_1\left(\vec{r}\right)\right|^2\right), \quad (7a)$$

where the wave functions are those of the Fermi level and the next level. Expanding in $H$, one finds the expression which, aside from the WFs and numerical coefficients, coincides with the expression for the total magnetic moment of the particle obtained from the partition function in Eq. (1). The same is true for odd electron particles. Therefore, we conclude that in a given system the Knight shift broadening should reflect the WF behavior, while on the average it should be a combined level statistics and WF effect. In particular, in the WD case the line shape of a single particle should follow the Poisson distribution of $\left|\psi\right|^2$ [7] (notice that since WFs of different energy states are distributed independently [7], the contributions from the electrons with opposite spins $\propto \left(\left|\psi_0\left(\vec{r}\right)\right|^2 - \left|\psi_1\left(\vec{r}\right)\right|^2\right)$ average out to zero in every particle). On the average, however, it can be shown that the WF distribution does not effect significantly the result for the line shape anticipated on the basis of level statistics alone. In the orthogonal case, for instance, the functional dependence of the line shape should follow closely that of Eq. (5b) in the limit $\varepsilon \ll 1$ (see Fig. 2). In classically regular systems, on the other hand, the WF behavior can be quite complex [7] and we will leave the discussion of the effect this might have on the Knight shift to future work.

9.     We will now discuss the relationship of the above results to other mechanisms and regimes for the Knight shift broadening. Firstly, Efetov and Prigodin [12] have considered the same problem. We believe, however, that their model, wherein the chemical potential is presumed fixed and the level broadening is smaller than the level spacing, is unrealistic. In the preceding paper [9], we discussed the susceptibility fluctuations in the opposite limit,

$\beta\delta \ll 1$. It should be emphasized that contrary to the wide-spread belief found in the small

particle literature [5], the discreteness of the spectrum is manifested even in this limit [9],

$$\left\langle \left(\chi - \langle\chi\rangle\right)^2 \right\rangle \sim \begin{array}{ll} \chi_P^2(\delta\beta), & \text{Poisson} \\ \chi_P^2(\delta\beta)^2, & \text{WD} \end{array} . \tag{8}$$

The analysis in Ref. [9] can be extended to higher moments ($f' \equiv \partial_\varepsilon f$),

$$\begin{aligned} &\left\langle \left(\chi - \langle\chi\rangle\right)^n \right\rangle = \\ &\mu_B^{2n} \int\int \left\langle \delta\rho(\varepsilon_1)\delta\rho(\varepsilon_2)\ldots\delta\rho(\varepsilon_n) \right\rangle f'\left(\varepsilon_1, \langle\mu\rangle\right) f'\left(\varepsilon_2, \langle\mu\rangle\right)\ldots f'\left(\varepsilon_n, \langle\mu\rangle\right) d\varepsilon_1 d\varepsilon_2 \end{aligned}, \tag{9}$$

where $\langle\chi\rangle \cong \chi_P$ [9]. It can be argued that in the considered limit the largest contribution to

Eq. (9) comes from pair-wise decoupling of the $n$-point level density correlation function.

Elementary combinatorics then yields a Gaussian distribution of the susceptibility. Thus, a

narrow Gaussian distribution centered around $\chi_P$ transforms to a broad distribution whose

tail extends to $\sim \chi_C$ as the temperature decreases below the mean level spacing. It might be

possible to test the susceptibility distribution for $\beta\delta \ll 1$ by measuring the line shape of the

Knight shift as well.

10     Previously, we have predicted [8] inhomogeneous broadening of the Knight shift whose origin is the large orbital currents in disordered metals. In the presence of spin-orbit scatterers, such currents can also couple to spin degrees of freedom [8]. In small particles, the local susceptibility fluctuations were predicted to be of the order of magnitude of the per unit volume Pauli (Landau) susceptibility. These effects should be compared to the Zeeman effects studied here. Clearly, for $\beta\delta \ll 1$ the orbital and spin-orbit effects will be larger (the former should be identifiable, however, in quasi-2D systems in an in-plane magnetic field).

At this point, we cannot definitively predict the magnitude of orbital and spin-orbit effects for $\beta\delta \gg 1$ since the derivation of Ref. [8] assumes a fixed chemical potential. We argued in Ref. [9] that such an assumption would definitely not yield considerable difference with the canonical ensemble only in the opposite limit, $\beta \delta \ll 1$. But since Ref. [8] predicts little temperature sensitivity of the fluctuations (once the length $L_T = \sqrt{\hbar D / T}$, where $D$ is the diffusion coefficient, becomes larger than the particle size), we can extrapolate its result to $\beta^{-1} \sim \delta$. Notice that the upper bound on the applicability of linear response is different for the orbital and spin mechanisms. For the orbital response, we refer to its discussion in Ref. [8] where it was shown that the magnetic moment of a particle can be as large as $\mu_B (k_F \ell)$, where $\ell$ is the electron mean-free-path and $k_F$ is length of the Fermi wave vector. For the spin response, the upper bound of the linear regime is $\mu_B H < \beta^{-1}$. Therefore, for $\beta \delta \gg 1$, the maximum magnetic moment is $\sim \chi_C H < \mu_B$; this estimate holds up to $\beta^{-1} \sim \delta$. Clearly, a more meaningful comparison requires a better understanding of orbital effects in the limit $\beta \delta \gg 1$.

11.     In conclusion, using the two- and three-level model, we determined the distribution function of the spin susceptibility for even- and odd-electron particles in the limit $\beta \delta \gg 1$. In the even case, the approximate form of the distributions for the Poisson and orthogonal level statistics are given by Eqs. (5). They can be characterized as being roughly power-law functions with integrable singularities for small susceptibilities and tails extending to $\sim \chi_C$ for large susceptibilities. The main difference between the two is in the prefactors $\varepsilon$ and $\varepsilon^2$ respectively, where $\varepsilon \equiv (\beta \delta)^{-1} \ll 1$. In the odd case, the distribution functions are narrowed further by additional factors of $\varepsilon$ and $\varepsilon^2$ respectively and are peaked at $\sim \chi_C$. Although in either case the distribution function narrows as the temperature is decreased, the tails are a distinct feature that should be experimentally identifiable. The larger broadening in the even case is a reversed effect relative to the effect expected from the particle size distribution [5].

12.     We enjoyed several illuminating discussions with B. Goodman. We used Theorist and Mathematica for Power Macintosh for some of the analytical calculations and plotting. This research was supported by the Army Research Office Grant Number DAAL03-93-G-0077.

Appendix I

Here we evaluate the moments of the susceptibility distribution function obtained in the two- and three-level approximations for a system with Poisson level statistics. Based on their comparison, we shall argue that the few-level model is less accurate for the tails of the distribution. The $n$'th moment (since $\langle \chi \rangle \sim \varepsilon$, it can be neglected) is given by,

$$\langle \chi^n \rangle = \int_0^\infty d\Delta \left( \frac{8t}{1+4t+t^2} \right)^n P\left(\Delta\right) = \varepsilon \int_0^1 dt \left( \frac{8t}{1+4t+t^2} \right)^n t^{\varepsilon-1} \ , \qquad \text{(AI-1)}$$

where we have used the notation of Eq. (2) and Eq. (3a). In the large $n$ limit, the multiplier $t^{\varepsilon-1}$ can obviously be neglected. Although we could not find a closed form for the integral, it is clear that it scales roughly as $\alpha^n$, where $\alpha > 1$ and is close to $\frac{4}{3}$, the maximum value of the susceptibility in the two-level model which is achieved at $t=1$ ($\Delta = 0$). This result is consistent with the evaluation of $\langle \chi^n \rangle$ with the distribution function (5a), which validates the approximation made in the derivation of the latter. Notice that in the orthogonal case one can use the first equality in (AI-1) and the distribution function (3b) to obtain $\langle \chi^n \rangle \sim \varepsilon^2$.

Using the three-level partition function from Halperin's review [5], we find that the lowest order correction to the two-level moment is given by,

$$\langle \chi^n \rangle_3 = \iint_{0\infty}^{\ 0\ 1} d\Delta d\Delta' \left( \frac{8t}{1+4t+t^2} \right)^n \left(1+t'\right)^n P\left(\Delta, \Delta'\right) = \langle \chi^n \rangle_2 \int_0^\infty d\Delta' \left(1+t'\right)^n P\left(\Delta'\right) \ , \quad \text{(AI-2)}$$

where we used $P\left(\Delta, \Delta'\right) = P\left(\Delta\right) P\left(\Delta'\right)$ for the Poisson ensemble [1]. By means of scaling the integration variable and series expansion we obtain,

$$\langle \chi^n \rangle_3 = \langle \chi^n \rangle_2 \varepsilon \int_0^\infty dx \left(1+e^{-x}\right)^n e^{-\varepsilon x} \cong \langle \chi^n \rangle_2 \left(1+\varepsilon \sum_{m=1}^n \binom{n}{m}\frac{1}{m}\right) . \qquad \text{(AI-3)}$$

The sum in (AI-3) can be evaluated in a closed form in terms of hypergeometric functions. For our purposes, however, it is sufficient to notice that in the large $n$ limit it is limited by $2^n / n$ and $2^n$ from below and above respectively [13]. This demonstrates that, for any $\varepsilon$, there will be a large correction to a high cumulant from the account of more levels near the Fermi surface. Therefore, we conclude that the tails of the distribution functions are not as well defined in the few-level model. This conclusion can be confirmed by a simple physical argument. Indeed, we notice that the maximum value of the susceptibility changes with the account of more levels. For instance, it is $\frac{4}{3}$ in the two- and $\frac{9}{5}$ in the three-level models. These values are found from $\Delta = 0$ and $\Delta = \Delta' = 0$ respectively. In other words, the tails are due to the systems with several levels within $k_B T$ of the Fermi level, which, in turn, indicates that a large number of levels must be taken into account for their characterization.

Appendix II

Here we detail a formal extension of our previous derivation [9], valid for $\beta\delta \ll 1$, to $\beta\delta \gg 1$. We also show that it is equivalent to the derivation of Ref. [12]. We begin with the expression for the $n$'th moment of the susceptibility,

$$\left\langle \chi^n \right\rangle = \left\langle \left[ \mu_B^2 \sum_k f'(\varepsilon_k) \right]^n \right\rangle \cong \frac{\mu_B^{2n} \beta^n}{2^{2n}} \sum_k \cosh^{-2n}\left( \frac{\varepsilon_k \beta}{2} \right),$$ (AII-1)

where we have used the diagonal approximation which gives the largest contribution in the limit $\beta\delta \gg 1$. Converting (A1) to an integral on energy with the level density $\delta^{-1}$, we find, upon some manipulations and integration [13],

$$\left\langle \chi^n \right\rangle = \frac{\mu_B^{2n} \beta^{n-1}}{2^{2n-1}\delta} \int_{-\infty}^{-\infty} dx \cosh^{-2n}(x) = \frac{(n-1)!}{2^{n-1}(2n-1)!!} \chi_C^{n-1} \chi_P.$$ (AII-2)

To obtain the local moments, one must also Poisson-average the $n$'th power of the squared WF, which yields an extra factor of $n!$. The latter is illustrated for the model considered in Ref. [12] wherein $\beta^{-1} = 0$ but a finite level broadening is assumed. From Eq. (4.7) in the second of Refs. [12], we obtain in the diagonal approximation and with the help of simple integration [13],

$$\begin{aligned} \left\langle \chi^n \right\rangle &= \frac{\mu_B^{2n}}{\delta} \int_{-\infty}^{\infty} d\varepsilon \int_0^{\infty} dv \left[ \frac{(\gamma/2\pi)v}{\varepsilon^2 + (\gamma/2)^2} \right]^n \exp(-v) \\ &= \frac{\mu_B^{2n}}{\delta}\left(\frac{1}{\gamma\pi}\right)^{n-1} \frac{n!(2n-3)!!}{(n-1)!} = \left(\frac{\mu_B^2}{\delta}\right)^n \left(\frac{1}{2\alpha}\right)^{n-1} \frac{n(2n-2)!}{(n-1)!}, \end{aligned}$$ (AII-3)

which coincides with Eq. (4.21). Moreover, the characteristic function formalism allows to restore the distribution function from the moments [14]. For instance, a trivial calculation using (AII-3) yields the following distribution function:

$$p(x) = \theta(x)\delta(x) + \left(\frac{\alpha}{8\pi x^3}\right)^{1/2}\exp\left(-\frac{\alpha x}{2}\right)(1+\alpha x), \qquad\qquad (AII\text{-}4)$$

where $\theta(x)$ is the step-function and $x = \chi/\chi_P$. The second term in Eq. (AII-4) obviously reproduces the limiting results given by Eqs. (4.16) and (4.18) in the second of Refs. [12], while its first term coincides with the limit $\alpha \to 0$ of Eq. (4.6). In other words, once the WF distribution is known [7], the derivation becomes transparent and the application of the non-linear $\sigma$-model is unnecessary, at least for $x > \alpha$. More importantly, we stress the formal nature of the above derivation in the limit $\beta\delta \gg 1$ since the first of equalities in Eq. (AII-1) is valid only in the opposite limit $\beta\delta \ll 1$. Furthermore, its results, given by Eqs. (AII-2) and (AII-3), are clearly statistics-independent, while the correct results (see text and Appendix I) clearly are. It helps, on the other hand, to establish a connection between our results for $\beta\delta \ll 1$ [9] and the results of a few-level model which is appropriate for $\beta\delta \gg 1$ and is a subject of this work.

Appendix III

Here we derive the susceptibility distribution function for the odd-electron particles with Poisson level statistics. Using the partition function in Halperin's review [5], we find,

$$
\begin{aligned}
p(\chi) &= \iint_0^\infty d\Delta \, d\Delta' \, \delta\Big(\chi - 1 + 8tt'\big(1 - (t+t')\big)\Big) P(\Delta, \Delta') \\
&= \varepsilon^2 \iint_0^1 dt \, dt' \, \delta\Big(\chi - 1 + 8tt'\big(1 - (t+t')\big)\Big)(tt')^{\varepsilon-1}
\end{aligned}
, \qquad \text{(AIII-1)}
$$

where we have used Eq. (3a) and the approximation $t, t' \ll 1$. Without the last term inside the delta-function, one recovers $p(\chi) = \delta(\chi - 1)$, the Curie susceptibility. The structure of the second integral is as follows. Without the account for the $(t + t')$ term, one would have $\varepsilon^2\big((\chi - 1)/8\big)^{\varepsilon-1}$ for $\chi > 1$, but the integral on second variable would yield a logarithmic divergence at zero. The account of this term, however, shows that the divergence is cut-off by $(\chi - 1)/8$. Consequently, we find,

$$
p(\chi) = -\varepsilon^2\big((\chi - 1)/8\big)^{\varepsilon-1} \ln\big((\chi - 1)/8\big) \qquad \text{(AIII-2)}
$$

which, up to the extra factor of $\varepsilon$, has virtually the same behavior at $1$ as the distribution function (5a) at zero. Notice that in the three-level approximation used here, the maximum susceptibility is $\dfrac{9}{5}$.

<u>Figure Captions</u>

Figure 1: $p\left(\chi\right)$ given by Eq. (5b) (solid line) and $p\left(\chi\right)$ obtained as a result of exact integration in Eq. (2) (dashed line) for $\varepsilon = 0.2$ .

Figure 2: $p\left(\chi\right)$ given by Eq. (5b) (solid line) and $\int_0^\infty dv \exp\left(-v\right) v^{-1} p\left(\frac{\chi}{v}\right)$ (dashed line) for $\varepsilon = 0.2$ .

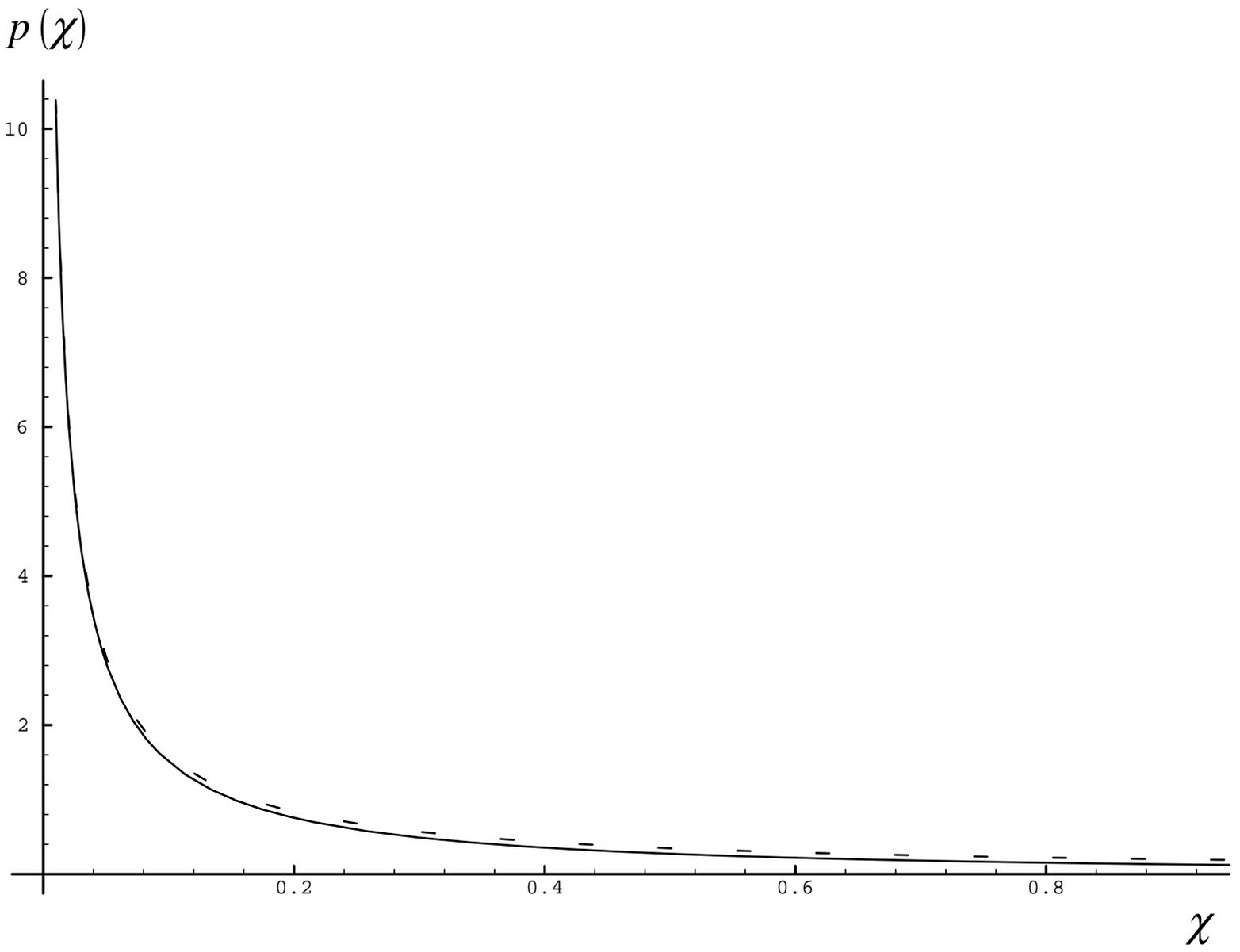

Figure 1

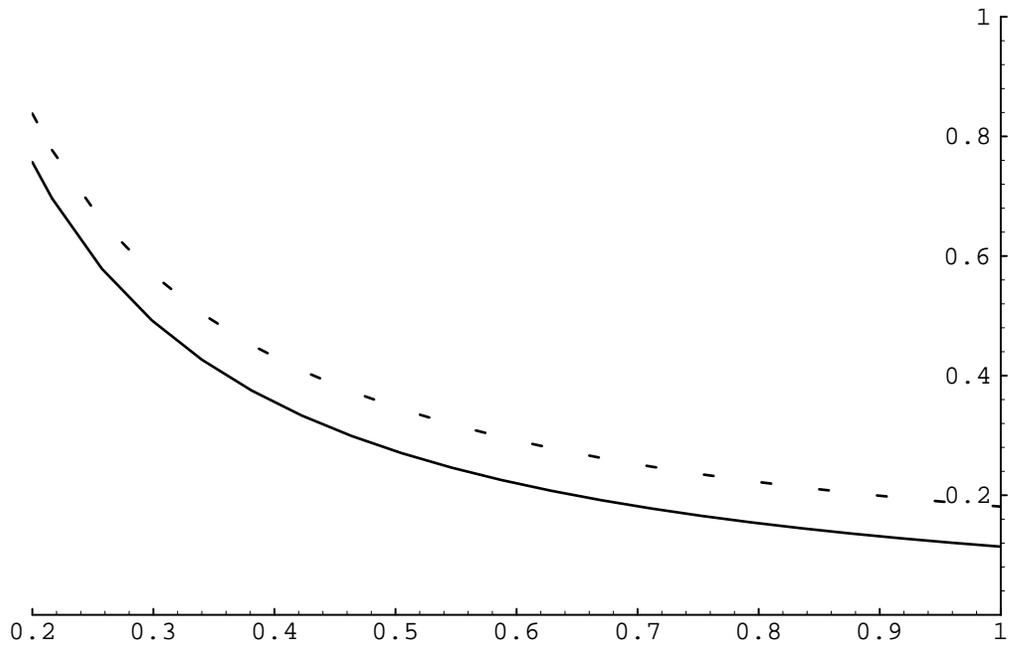

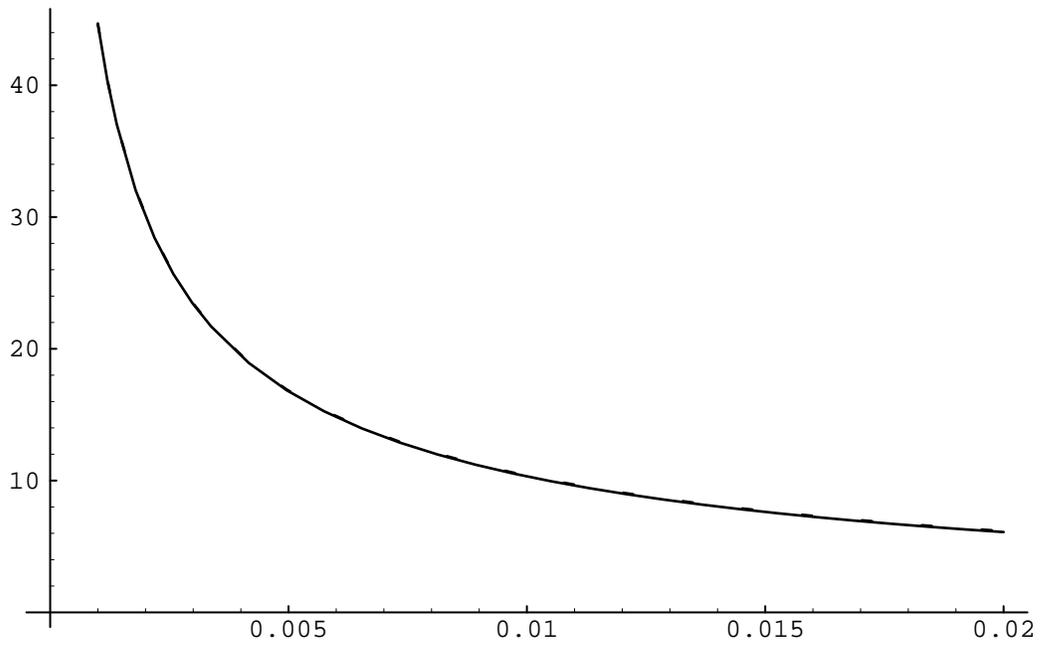

Figure 1 (inserts)

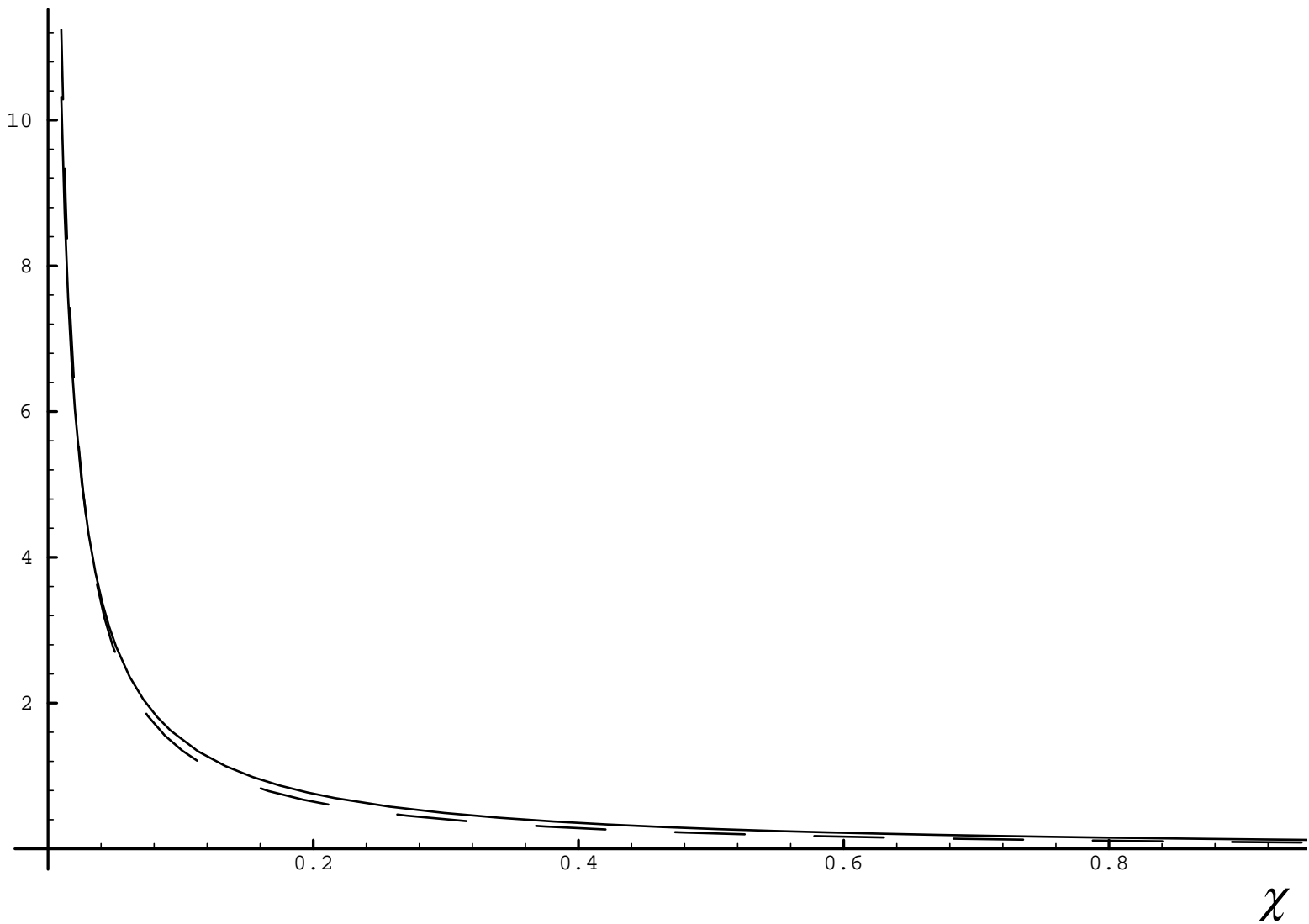

$$p(\chi), \qquad \int_0^\infty dv \exp(-v) v^{-1} p\left(\frac{\chi}{v}\right)$$

Figure 2

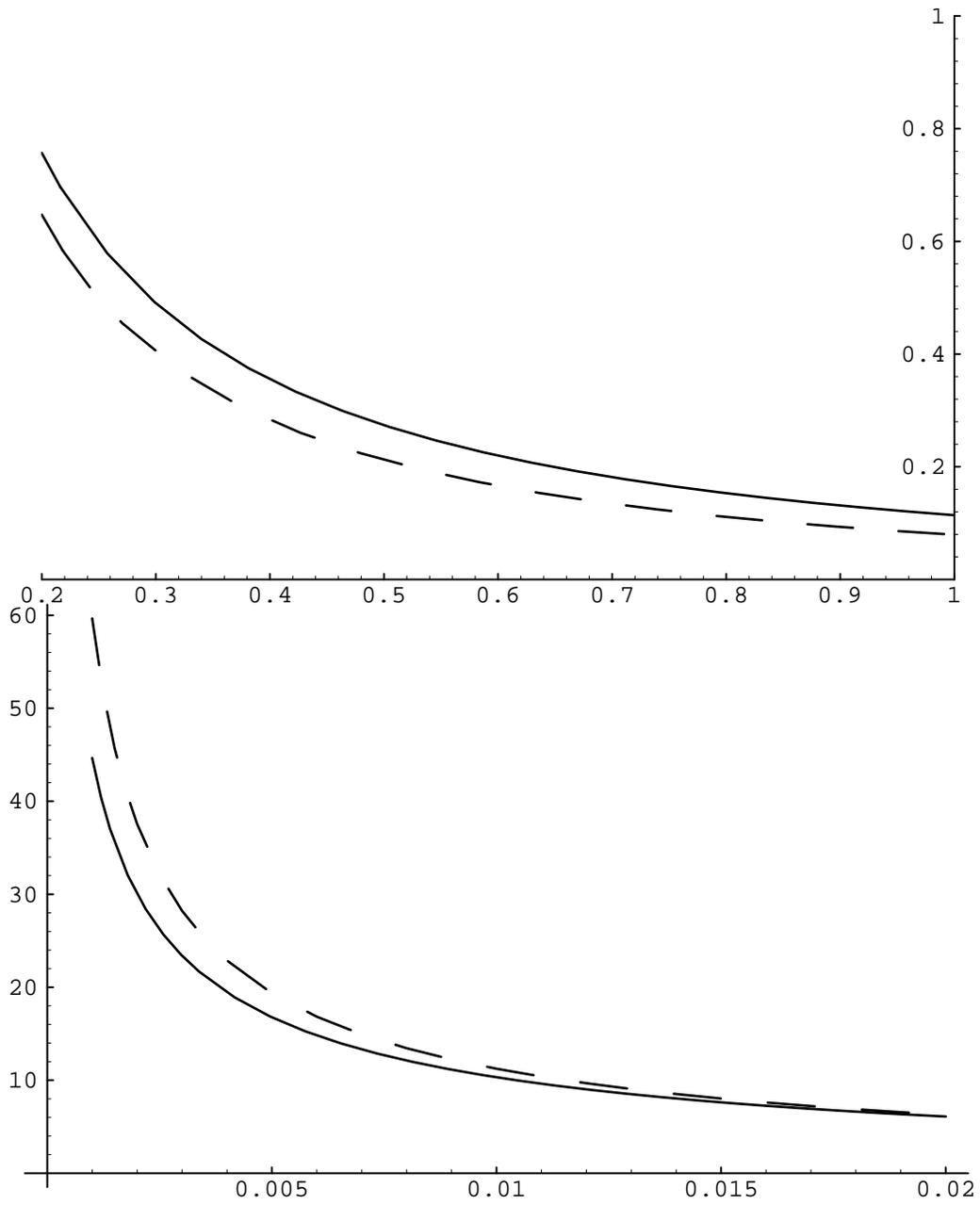

Figure 2 (inserts)